\documentclass[12pt,a4paper]{article}

\usepackage[british]{babel}

\usepackage[a4paper,top=2cm,bottom=2cm,left=2.5cm,right=2.5cm,marginparwidth=1.75cm]{geometry}

\usepackage{amsmath}
\usepackage{graphicx}
\usepackage[colorlinks=true, allcolors=blue]{hyperref}
\usepackage{hyperref}
\usepackage[title]{appendix}
\usepackage{mathrsfs}
\usepackage{amsfonts}
\usepackage{booktabs} 
\usepackage{caption}  
\usepackage{threeparttable} 
\usepackage{algorithm}
\usepackage{algorithmicx}
\usepackage{algpseudocode}
\usepackage{listings}
\usepackage{enumitem}
\usepackage{chngcntr}
\usepackage{booktabs}
\usepackage{lipsum}
\usepackage{subcaption}
\usepackage{authblk}
\usepackage[T1]{fontenc}    
\usepackage{csquotes}       
\usepackage{diagbox}
\usepackage{geometry}
\usepackage{tocloft}
\usepackage{graphicx}
\usepackage{fancyhdr}
\usepackage{amsmath}
\usepackage{listings}
\usepackage{xcolor}
\usepackage{hyperref}
\usepackage{hyperref}
\hypersetup{
    colorlinks=true,
    linkcolor=blue,
    citecolor=blue,
    urlcolor=blue
}

\lstdefinestyle{mystyle}{
    language=Python,
    basicstyle=\small\ttfamily,
    keywordstyle=\color{blue},
    commentstyle=\color{green!40!black},
    stringstyle=\color{red},
    numbers=left,
    numberstyle=\tiny,
    numbersep=5pt,
    frame=single,
    breaklines=true,
    breakatwhitespace=true,
    tabsize=4,
    captionpos=b,
}
\usepackage{setspace}
\usepackage{titlesec}
\titleformat{\section} 
  {\normalfont\Large\bfseries}{\thesection.}{1em}{}
  
\usepackage{lineno} 

\usepackage{float}   
\usepackage{caption} 


\usepackage[nottoc]{tocbibind}

\title{Knot-detection algorithm to measure viscosity in three-dimensional MHD plasmas}
\author{Ratul Chakraborty$^\dagger$}
\author{Rupak Mukherjee$^\ddagger$}
\affil{Department of Physics, Sikkim University, Gangtok, Sikkim, India}

\date{\today}

\begin{document}

\maketitle

\begin{abstract}
This project explores the mathematical study of knots and links in topology, focusing on differentiating between the two-component Unlink and the Hopf Link using a computational tool named LINKAGE. LINKAGE employs the linking number, calculated through Barycentric Equations, Matrix Algebra, and basic topological principles, to quantify the degree of linking between two closed curves in three-dimensional space. This approach not only distinguishes between different knot structures but also has applications in understanding complex systems such as magnetic field lines in plasma physics. Additionally, this project includes an example where multiple interlinked loops were analyzed over different time stamps using the LINKAGE algorithm. By observing how these links break and evolve, the algorithm demonstrates its ability to track changes in the topological properties of the system. This dynamic analysis shows the versatility of the tool in studying evolving systems, where the topology of the components can change, providing valuable information about the underlying physical processes driving these changes.
\end{abstract}


\vspace{0.5cm}

\noindent
{\bf {$^\dagger$email:}} ratulchakraborty2016@gmail.com\\
{\bf {$^\ddagger$email:}} rmukherjee@cus.ac.in

\newpage

\tableofcontents

\section{Introduction}

\ Knots and links are common occurrences in our daily lives, whether it's tying shoelaces or securing a package. However, beyond these practical uses, knots and links take on a more abstract yet profound form in the realm of topology, where they are studied as mathematical objects with intriguing properties. These objects play a vital role in a variety of scientific fields, particularly in understanding complex systems in physics. In this work, we focus on to develop an algorithm to distinguish two fundamental structures in knot theory: the two-component Unlink and the Hopf Link. We lay out a potential use of our work in the domain of plasma physics.

We know, plasmas governed by the ideal magnetohydrodynamic (MHD) equations show a strict conservation of helicity. However, in presence of tiny amount of viscosity, plasmas can relax and form a steady-state at the late-time evolution. For an ideal gas system, this steady state can be derived by simply extremising the free energy of the system defined as $F = U - TS$, where, $U$, $T$ and $S$ are the internal energy, absolute temperature and entropy of the gas respectively. Such extremisation methods of the general steady states, arising out of such arguments, can shed light on understanding the underlying physics issues in a global way. For a near ideal plasma one can follow the similar arguments and derive one such extremisation principle \cite{woltjer:1958}. 

L. Woltjer in 1954 constructed the free energy using magnetic helicity ($H_m = \int \vec{A} \cdot \vec{B} ~dV $) and magnetic energy ($E_m = \int B^2 ~ dV$ and $\vec{B} \cdot \hat{n} = 0$) and extremised the same to yield $\vec{\nabla} \times \vec{B} = \alpha (x) \vec{B}$ subject to $\vec{B} \cdot \vec{\nabla} \alpha = 0$ on the ``walls'', where, $\alpha = \alpha (x)$ represents the rigidity of the large number of volume ($H_m = \int\limits_V \vec{A} \cdot \vec{B} ~ dV$ where, $V$ is the nested volume). Later it was found that, in presence of weak dissipation the local helicity constants in the bulk can break, leaving only one on the conducting surface\cite{taylor:1974}. Thus, $\alpha (x) = \alpha_0 = $ constant and hence, $\vec{\nabla} \times \vec{B} = \alpha_0 \vec{B}$ \cite{taylor:1974, taylor:1986}. Further modification of this model has been achieved in MRxMHD by Hudson {\it et. al.} \cite{hudson:2012} and the references therein.

In short, the final states of near ideal plasmas governed by MHD equations, with perfectly conducting boundaries, the steady-state will be a force-free Taylor-Woltjer state defined as $\vec{\nabla} \times \vec{B} = \alpha_0 \vec{B}$, where $\alpha_0$ is a constant and $\vec{B}$ is the magnetic field defined over a volume $V$. Such relaxed states under force-free condition are called Beltrami-type states in Cartesian geometry. Reversed-Field-Pinch devices - a device to study laboratory fusion via magnetised plasmas work exploiting such ideas. 

Much later, Mahajan and Yoshida showed that such relaxation model works in presence of plasma flows ($\vec{u}$) as well. They defined a generalised helicity $H_G = \int \left( \vec{u} + \vec{A} \right) \cdot \left( \vec{\omega} + \vec{B} \right) ~dV$ and Energy $W = \int \left( u^2 + B^2 \right) ~dV$. The relaxed steady-state was found to be a `Double Beltrami' flow \cite{mahajan:1998}. A further extension of the study showed that constant magnetic helicity ($H_m$) leads to a `Triple Beltrami' flow \cite{dasgupta:1998}. 

All these models so far relied on the fact that plasma relaxation takes place in presence of even a tiny viscosity, which only acts at the small scales of the system. But later, arbitrary scale relaxation model was proposed by H. Qin {\it et. al.}\cite{qin:2012}. Mukherjee {\it et. al.} \cite{mukherjee:2018b} showed that the relaxation spectra is dominantly controlled by $k \textgreater 10$ modes present in the system, thereby supporting the multiscale relaxation process suggested by H Qin. In this work, Mukherjee {\it et. al.} \cite{mukherjee:2018b} upgraded a compressible pseudo-spectral 3D MHD solver \cite{mukherjee:2018a, rupak_mukherjee_2021_4682188, mukherjee:2019c, mukherjee:2019a, mukherjee:2019b, mukherjee:2019d, mukherjee:2018c, biswas:2021} to evolve Arnold-Beltrami-Childress flow in a three dimensional Cartesian geometry using a conducting or periodic as well as mixed boundary condition. The plasma was `suddenly' relaxed to a double volume and all the wave numbers contributing to this relaxation process was measured. The results obtained in this study has been found to confirm the multiscale relaxation process.

    Here we summarise the above discussion for the purpose of this work. Consider a medium with zero viscosity, where two magnetic field lines are linked together. In such a scenario, these lines will remain linked indefinitely, since magnetic field lines cannot cross or break under ideal conditions. This characteristic of linked magnetic fields gives us a measure of helicity. In a non-viscous medium, helicity remains conserved over time, as the linked structures cannot change. However, when viscosity is introduced, magnetic field lines tend to reconnect or break, leading to changes in the medium's helicity. The rate of change in helicity becomes proportional to the viscosity of the medium, making helicity a valuable metric for measuring viscosity in such systems.
    
    A tool that can measure the variation in helicity over time would thus be able to quantify theviscosity of the medium. For this, the tool must be capable of distinguishing between a configuration where the magnetic field lines are linked and one where the links have been broken over time. The simplest example of a linked magnetic field structure is the Hopf Link, which consists of two interlinked loops. If this structure breaks apart, it results in two separate, unlinked rings, known as the Unlink. Detecting such a transition is crucial for understanding how helicity evolves in viscous media.
\begin{figure}[h]
    \begin{minipage}{0.35\textwidth}
        \centering
        \includegraphics[width=\textwidth]{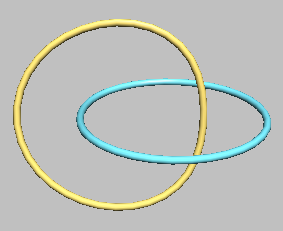}
        \caption{Hopf ring}
        \label{fig:image1}
    \end{minipage}\hfill
    \begin{minipage}{0.355\textwidth}
        \centering
        \includegraphics[width=\textwidth]{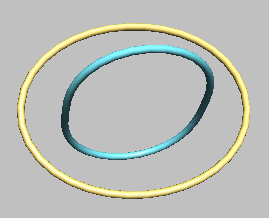}
        \caption{Unlink}
        \label{fig:image2}
    \end{minipage}
\end{figure}

    To address this challenge, there has been several attempts to develop such numerical framework \cite{arai:2013, qu:2021, fuller:1978, jacobson:2013, edelsbrunner:2003, bernstein:2013, van:2005, van:2006}. We describe the development of an open-source computational tool called LINKAGE. Our in-house developed software, LINKAGE utilizes a topological concept known as the linking number, a numerical value that quantifies the degree of linking between two closed loops in three-dimensional space. By applying Barycentric Equations for Lines and Planes, matrix algebra, and foundational principles of topology, LINKAGE constructs an efficient algorithm capable of calculating the linking number for any given pair of loops. Through this process, subtle differences between the Unlink and the Hopf Link are identified and quantified, much like how DNA fingerprinting distinguishes between individuals based on genetic markers.
    
    At the end this work includes a dynamic example where multiple interlinked loops were analyzed over different time stamps using the LINKAGE algorithm. By tracking how these links break and evolve, the algorithm showcases its ability to monitor changes in the topological properties of a system. This dynamic analysis underscores the versatility of LINKAGE in studying evolving systems, providing valuable information about the physical processes driving these changes, such as the visocsity of the medium or the rate of magnetic reconnection.

\section{Linking Number}
Linking number, a topological invariant for a given link, represents the linking of two knots or links in three-dimensional space. Although various links can have same linking number but if two links have different linking number then it is for sure that the links are different. Linking number of Hopf link is $1$ and linking number of Unlink is $0$ and thus we can say they are different. We will adopt a surface-centric approach for computing this linking number. But before delving into that, it is essential to understand Seifert Surfaces, which are orientable surfaces associated with knots or links, because it is the foundation of our algorithm.

\section{Seifert Surfaces}

In simple terms a surface like that of a regular strip with a sense of top and bottom is called an orientable surface. A German mathematician, Herbert Seifert, provided a valuable algorithm known as Seifert's Algorithm. In this algorithm, he proved that for any knot or link, one can construct an orientable surface with the knot or link as its surface boundary. These surfaces are named after him and are called Seifert's Surfaces.

Look at the images shown below. In the first image, Figure \ref{subfig:borromean}, there is an orientable surface with the Borromean ring as its boundary. In the second image, Figure \ref{subfig:hopf}, there is an orientable surface with the Hopf link as its boundary. In the last image, Figure \ref{subfig:unknot}, there are two sides of an orientable disk surface with the unknot as its boundary. Seifert has proven that for any given link or knot, one can construct such orientable surfaces with the knot or link as the surface boundary.

\begin{figure}[h]
    \centering
    \begin{minipage}{0.288\textwidth}
        \centering
        \includegraphics[width=0.8\textwidth]{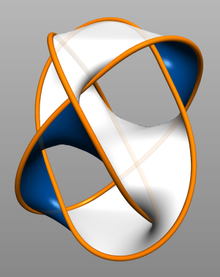}
        \caption{The Seifert surface of the Borromean ring with white and blue representing the two sides of the orientable surface. The orange band is the Borromean ring.}
        \label{subfig:borromean}
    \end{minipage}\hfill
    \begin{minipage}{0.28\textwidth}
        \centering
        \includegraphics[width=0.8\textwidth]{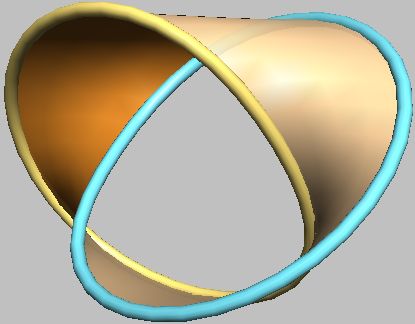}
        \caption{The Seifert surface of the Hopf link with buttermilk and brown representing the two sides of the orientable surface. The yellow and the blue rings constitute the Hopf ring.}
        \label{subfig:hopf}
    \end{minipage}\hfill
    \begin{minipage}{0.28\textwidth}
        \centering
        \includegraphics[width=0.8\textwidth]{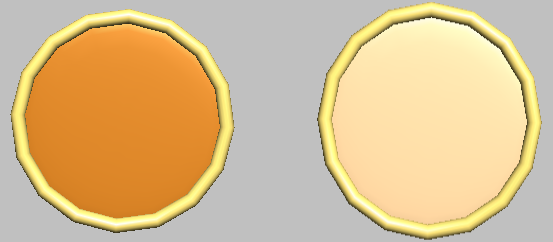}
        \caption{The Seifert surface of a single unknot. The two sides of the orientable surface are shown separately. The yellow ring constitutes the unknot.}
        \label{subfig:unknot}
    \end{minipage}
    
    \label{fig:seifert_surfaces}
\end{figure}
\section{Linking Number of Hopf Link}

Let's consider the case of a Hopf link. Given a Hopf link, we can have four different orientation versions of it, as shown in Figure 6 below.

\begin{figure}[h]
    \centering
    \includegraphics[width=0.3\textwidth]{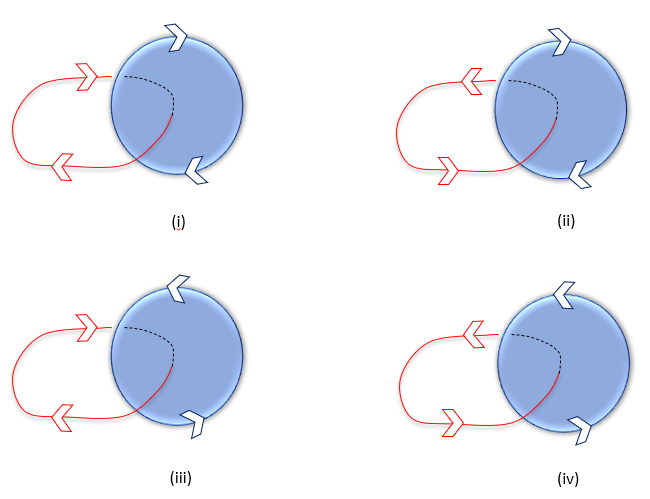}
    \caption{{Four possible orientations of the hopf link}}
    \label{fig:enter-label}
\end{figure}
We choose the face facing towards us as the top side and the face behind the screen as the bottom side for the Seifert surface of the blue loop. With this choice, in Figure 6.i and 6.iii, the red loop crosses the blue Seifert surface of the blue loop from bottom to top. There is only one crossing, and since the nature of these crossings is bottom to top we assign the value of $+1$ for both of them. Conversely, in Figure 6.ii and 6.iv, the red loop crosses the blue Seifert surface from top to bottom. Since the nature of the crossings is top to bottom, we assign the value of $-1$ for both of them. If we had chosen the back screen as the top side, the crossing value for Figure 6.i and 6.iii would have been $-1$ and $+1$ for Figure 6.ii and 6.iv. Regardless of the choice of orientation for the surface or the loop, the modulus in all cases is $1$. Thus, the linking number of a Hopf link is $1$. Aditionally we can also deform the Hopf link as per our will. However, all the deformations are homeomorphic to each other, and the linking number is always $1$.

\section{Linking Number of Unlink}
Let's attempt to implement the above perspective for the case of an unlink. First, consider the case of an unlink. Figure 7. 
 
\begin{figure}[h]
    \centering
    \includegraphics[width=0.6\textwidth]{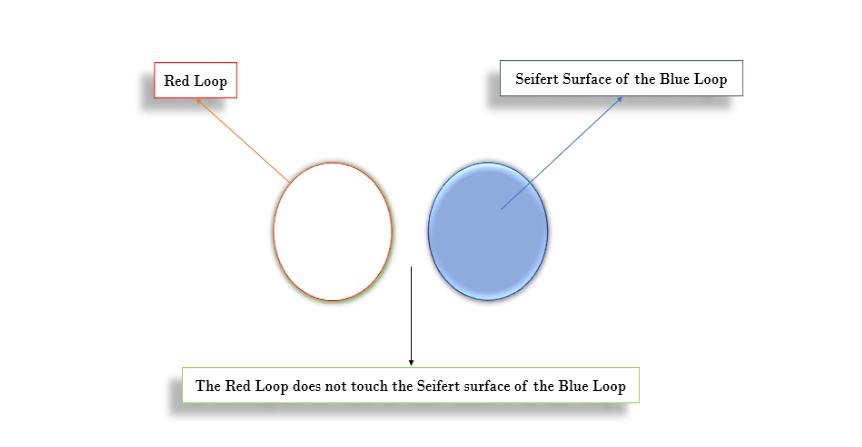}
    \caption{{Linking Number 0 for an Unlink}}
    \label{fig:enter-label}
\end{figure}
Since the red loop does not touch the Seifert surface of the blue loop, we assign the linking number $0$, meaning no crossings, for Unlink. The linking number remains $0$ for any deformations until and unless it crosses the blue surface. Some of the allowed deformations are illustrated below in Figure 8.

\begin{figure}[h]
    \centering
    \includegraphics[width=0.4\textwidth]{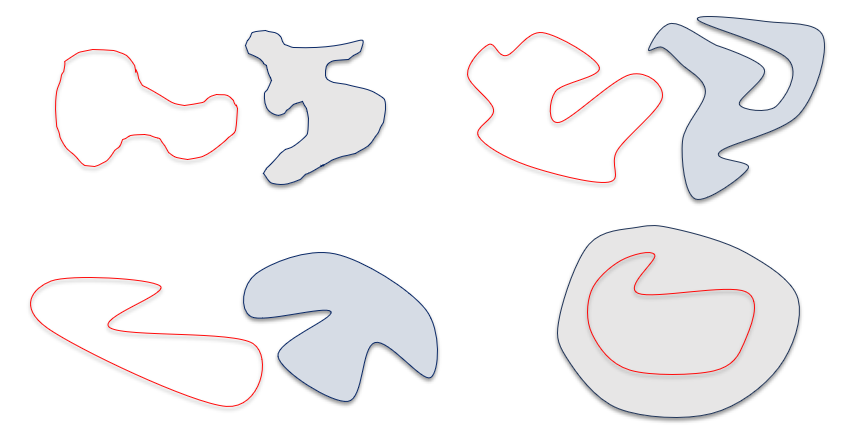}
    \caption{{Various ways in which we can deform the individual loops of the unlink}}
    \label{fig:enter-label}
\end{figure}
 
Thus based on the value of this linking number we can insist that a Hopf link is different from an Unlink, topologically.

\section{Numerical Computation of Linking Number}

\ To implement the above idea in a machine language we can ask ourselves a few important questions. These questions are discussed below one by one.

\subsection{Should we keep our loops (both red and blue) continuous or should we discretize them?}

If we want to keep the loops continuous, then we have to feed the algorithm the exact equations of the loops every time we compute. Two differently deformed unlinks will have different equations, and feeding the exact equations is easier for completely circular loops but becomes immensely difficult for much more complex deformations of the same. Ensuring an input for the algorithm becomes difficult this way, and thus, we should avoid relying on a continuous approach. Well then, how does discretization help in this matter?. What we mean by discretization of the loop is having a certain number of discrete points (data points) on the continuous loop and then joining each point with a straight line. The Figures 9 and 10 below illustrate an example of a discretized unlink and a discretized Hopf link.\

We provide the input link as two lists: one for the red loop and one for the blue loop. Each list contains a fixed number of 3D data points as tuples.

\begin{figure}[h]
    \begin{minipage}{0.45\textwidth}
        \centering
        \includegraphics[width=\textwidth]{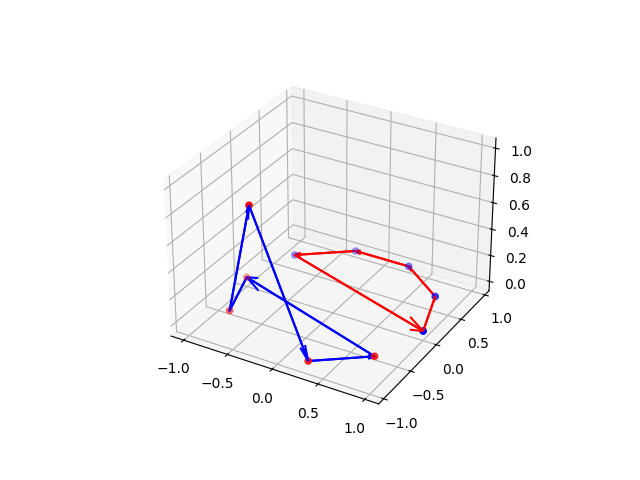}
        \caption{Discretized Unlink}
        \label{fig:image1}
    \end{minipage}\hfill
    \begin{minipage}{0.45\textwidth}
        \centering
        \includegraphics[width=\textwidth]{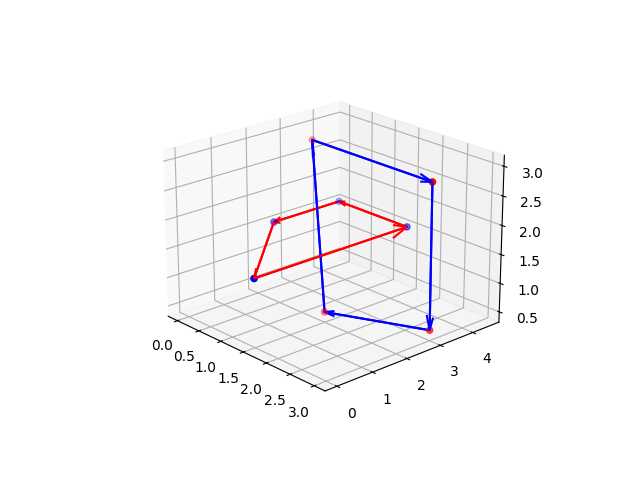}
        \caption{Discretized Hopf link}
        \label{fig:image2}
    \end{minipage}
\end{figure}

\subsection{How do we obtain the Seifert surface associated with the blue loop?}

\begin{algorithm}[!ht]
\caption{Constructing the Seifert Surface of the Blue Loop}
\begin{algorithmic}[1]
\Require Points set \texttt{points1} and 3D axis \texttt{ax}
\Ensure Seifert surface for the blue loop

\State Initialize an empty list for triangles: \texttt{triangles} $\Leftarrow$ \texttt{[]}
\For{$i \gets 1$ to $\texttt{len(points1)} - 2$}
    \State $\texttt{triangle} \gets (\texttt{points1}[0], \texttt{points1}[i], \texttt{points1}[i+1])$
    \State Append \texttt{triangle} to \texttt{triangles}
\EndFor
\State Create a 3D polygon collection \texttt{loop} from \texttt{triangles} using $\texttt{Poly3DCollection}$ with edge color 'k', line widths 1, and transparency 0.5
\end{algorithmic}
\end{algorithm}

To obtain the Seifert surface of the blue loop from it's data points, we apply triangulation. In Algorithm 1 we designate the first point, $point1[0]$, as the base point and consider the points $point1[i]$ and $point1[i+1]$ as the other two points. We construct triangles, where $i$ varies from $1$ to until $[len(points1)-1]$, up to $[len(points1)-2]$. These triangles are then appended to the list $triangles$. Subsequently, the code plots the triangles, along with their respective visualized surfaces. An example for Hopf link is depicted below in Figure 11. The Seifert surface obtained is not a smooth surface, rather it is a discrete triangulated approximation of the same. The surface tends to become smooth when we increase the number of data points on the loop.

\begin{figure}[h]
    \centering
    \includegraphics[width=0.7\textwidth]{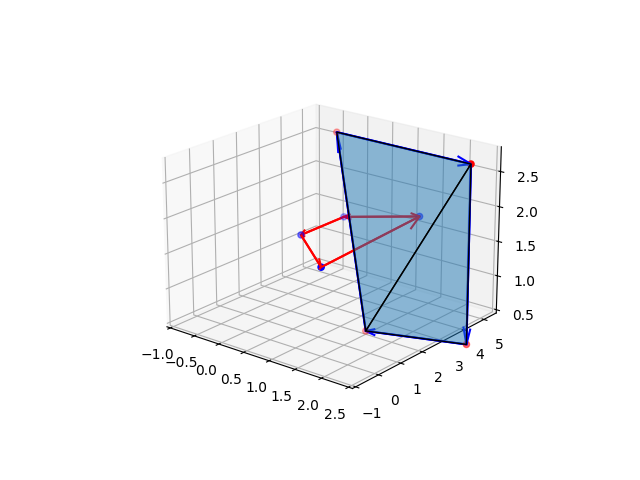}
    \caption{{The triangulated Seifert surface associated with the blue loop}}
    \label{fig:enter-label}
\end{figure}

\subsection{Crossing the Seifert surface: top to bottom and bottom to  top}

\begin{algorithm}[!ht]
\caption{Barycentric Plane and Line Equations for Finding Crossing Points}
\begin{algorithmic}[1]
\Require \texttt{points2}, \texttt{triangles}, symbolic variables \texttt{t}, \texttt{u}, \texttt{v}
\Ensure List of crossing points with orientation

\State Define \texttt{Line\_equation(point0, point1)}:
\State \texttt{return np.array(point0) - (np.array(point1) - np.array(point0)) * t}

\State \textbf{Print} 'Parametric equations of red vectors:'
\State Initialize \texttt{Red\_Vectors} as empty list
\For{$i \gets 0$ to $\texttt{len(points2) - 1}$}
    \State \texttt{Red\_Vectors.append(Line\_equation(points2[i], points2[(i + 1) \% \texttt{len(points2)]}))}
\EndFor
\State \textbf{Print} \texttt{Red\_Vectors}

\State Define \texttt{Plane\_equation(vertices0, vertices1, vertices2)}:
\State \texttt{return plane equation}

\State \textbf{Print} 'Parametric equations of blue triangles:'
\State Initialize \texttt{Blue\_Triangles} as empty list
\For{each \texttt{tri} in \texttt{triangles}}
    \State \texttt{Blue\_Triangles.append(Plane\_equation(tri[0], tri[1], tri[2]))}
\EndFor
\State \textbf{Print} \texttt{Blue\_Triangles}

\State Define \texttt{Solution(La, Lb, P0, P1, P2)}:
\State \texttt{Solve the system of linear equations}
\State \textbf{Print} \texttt{solution}
\If{solution is valid}
    \State \textbf{Print} \texttt{solution}
    \State Compute orientation
    \If{0 $\leq$ angle < 90} or {270 < angle $\leq$ 360} 
        \State \texttt{Linking\_List.append(1)}
    \ElsIf{90 < angle < 270}
        \State \texttt{Linking\_List.append(-1)}
    \EndIf
    \State \textbf{Print} \texttt{Linking\_List}
\Else
    \State \textbf{Print} 'Solution does not exist'
\EndIf

\end{algorithmic}
\end{algorithm}

The algorithm 2 uses a barycentric approach to calculate intersections between red vectors and blue triangular surfaces. First, it computes the barycentric line equations for the red vectors with the function Line\_equation and the plane equations for the blue triangles using Plane\_equation. For each triangular surface, the code solves the line equation for all red vectors. If the solution lies inside the triangle, it prints the result; otherwise, it prints "Solution does not exist." This process repeats for all triangular surfaces in a loop to compute crossing points. Vectors lying on the same plane as a triangle, whether inside or outside, are not considered crossing, so the code returns "Solution does not exist" in such cases. Next. we utilize the direction of the gradient vector to the surface at the point of intersection, to determine the direction of crossing.

\begin{algorithm}[!ht]
\caption{Orientation of the Seifert Surface}
\begin{algorithmic}[1]
\Require Values \texttt{u\_value}, \texttt{v\_value}, vectors \texttt{P0}, \texttt{P1}, \texttt{P2}, \texttt{La}, \texttt{Lb}
\Ensure List of orientations

\State Define \texttt{Gradient\_Vector(u\_value, v\_value)}:
\State \texttt{P01} $\gets$ \texttt{P1 - P0}
\State \texttt{P02} $\gets$ \texttt{P2 - P0}
\State \texttt{normal\_vector} $\gets$ \texttt{np.cross(P01, P02)}
\State \texttt{solution\_vector} $\gets$ \texttt{np.array(Lb) - np.array(La)}
\State \texttt{angle} $\gets$ \texttt{np.degrees(np.arccos(np.dot(normal\_vector, solution\_vector) / (np.linalg.norm(normal\_vector) * np.linalg.norm(solution\_vector))))}

\State \textbf{Print} \texttt{angle}

\If{$0 \leq \texttt{angle} < 90$ or $270 < \texttt{angle} \leq 360$}
    \State \texttt{Linking\_List.append(1)}
\Else
    \State \texttt{Linking\_List.append(-1)}
\EndIf

\State \textbf{Print} \texttt{Linking\_List}

\State \textbf{Print} \texttt{Gradient\_Vector(solution[u], solution[v])}

\end{algorithmic}
\end{algorithm}

We compute the angle between the normal vector of the Seifert surface and the vector that crosses the Seifert surface to determine if the crossing is positive or negative. This helps in understanding the orientation of the Seifert surface. The angle is calculated using the dot product and normalized vectors, with results printed to show the orientation in Linking List.

If the angle falls within the range $[0,90)$ or $(270,360]$, the red vector is considered to have crossed the surface from the bottom to the top side. Conversely, if the angle is in the range $(90,270)$, the red vector is considered to have crossed the surface from the top to the bottom side. This is how the orientation is determined. The code then prints the value of the angle.

\subsection{How do we compute the linking number for both the links at the same time and then compare them?}

If the crossing is from bottom to top, we append \(+1\) to our previously initiated empty list \textit{Linking\_List}, and \(-1\) if the crossing is from top to bottom. Once all the crossings have registered themselves in the \textit{Linking\_List}, we compute the sum of all the \(+1\)'s and \(-1\)'s, storing the result in the variable \textit{Linking\_Number}. We then take the absolute value of this variable using the \textit{abs} function to obtain the linking number for the link. In our case, since we are dealing with a Hopf link, the value of the linking number that the code returns is \(1\). For any unlink the value returned will be $0$. We encapsulate all of the above code within the function \textit{LINKING\_NUMBER}. When provided with the information of a link in the form of data points, the function returns the linking number for the link. This function, along with a code segment that compares the linking numbers of two links, serves as our final algorithmic function.

\section{Comparison between the two links}

\subsection{Comparison between two different Hopf links(Fig.12,Fig.13)}

\begin{figure}[h]
    \begin{minipage}{0.45\textwidth}
        \centering
        \includegraphics[width=\textwidth]{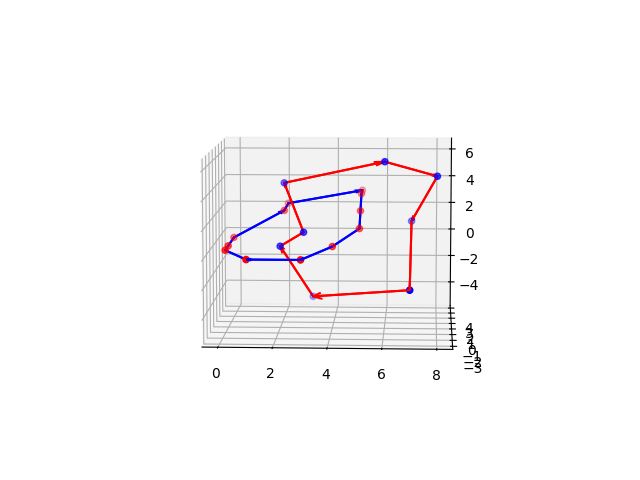}
        \caption{Hopf link 1}
        \label{fig:image1}
    \end{minipage}\hfill
    \begin{minipage}{0.45\textwidth}
        \centering
        \includegraphics[width=\textwidth]{disc_hpolk.png}
        \caption{Hopf link 2}
        \label{fig:image2}
    \end{minipage}
\end{figure}

\begin{lstlisting}
# Output
THE GIVEN LINKS ARE  SAME.
\end{lstlisting}

\subsection{Comparison between two different Unlinks(Fig.14,Fig.15)}
\begin{figure}[h]
    \begin{minipage}{0.45\textwidth}
        \centering
        \includegraphics[width=\textwidth]{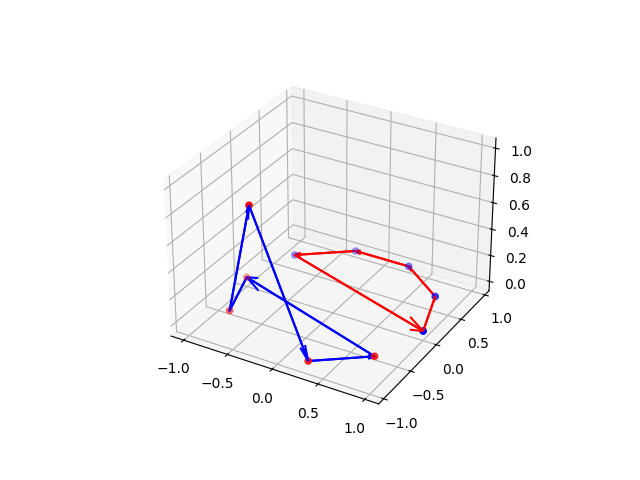}
        \caption{Unlink1}
        \label{fig:image1}
    \end{minipage}\hfill
    \begin{minipage}{0.45\textwidth}
        \centering
        \includegraphics[width=\textwidth]{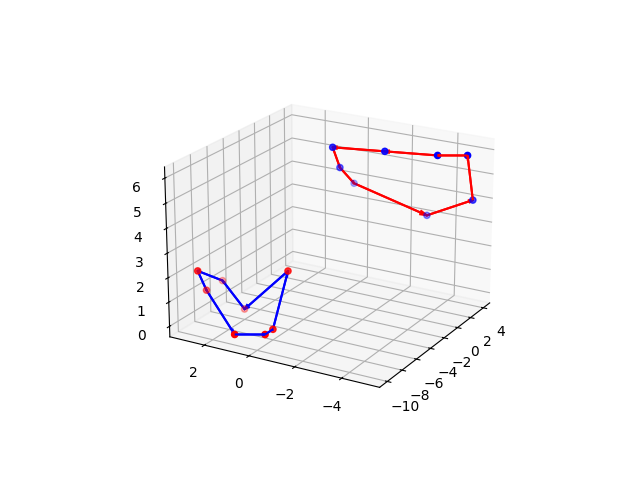}
        \caption{Unlink 2}
        \label{fig:image2}
    \end{minipage}
\end{figure}

\begin{lstlisting}
# Output
THE GIVEN LINKS ARE  SAME.
\end{lstlisting}

\subsection{Comparison between a Hopf link and an Unlink(Fig.16,Fig.17)}

\begin{lstlisting}
# Output
THE GIVEN LINKS ARE DIFFERENT.
\end{lstlisting}

\begin{figure}[h]
    \centering
    \begin{minipage}{0.45\textwidth}
        \centering
        \includegraphics[width=\textwidth]{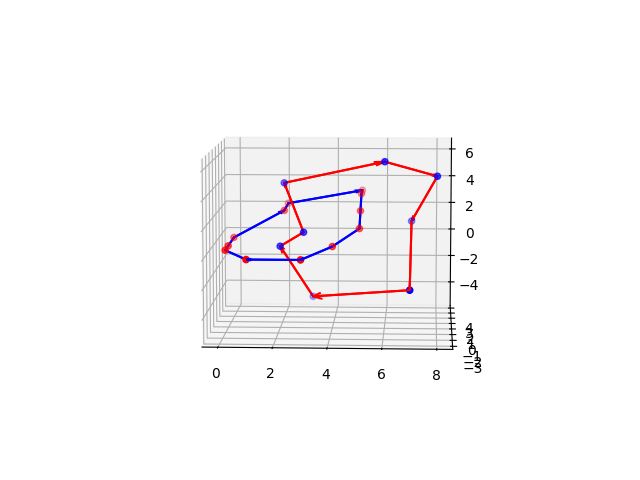}
        \caption{Hopf link}
        \label{fig:image1}
    \end{minipage}\hfill
    \begin{minipage}{0.45\textwidth}
        \centering
        \includegraphics[width=\textwidth]{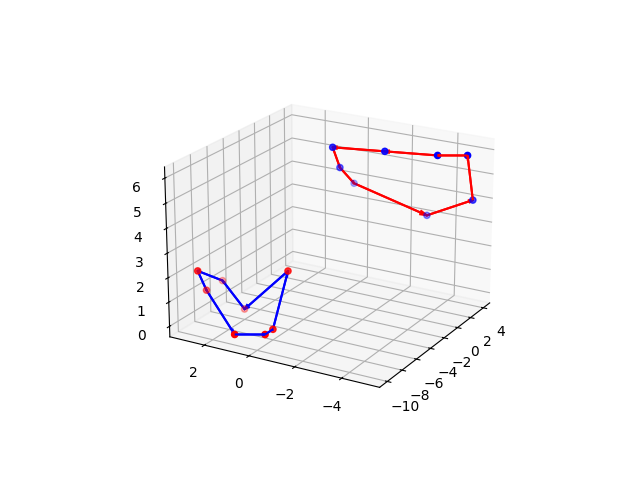}
        \caption{Unlink}
        \label{fig:image2}
    \end{minipage}
\end{figure}

\section{Study of Multiple Interlinked Loops Using LINKAGE}

In this section, we extend the study from two component structures to multiple component structures. We explore the potential of such linking number algorithms for studying structures involving multiple interlinked loops. The complex structure we consider is shown in the Figure 18 below.

\begin{figure}[h]
    \begin{minipage}{0.35\textwidth}
        \centering
        \includegraphics[width=\textwidth]{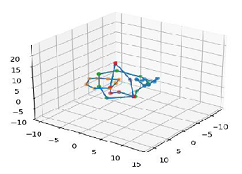}
        \caption{Four loops interlinked with each other.}
        \label{fig:image1}
    \end{minipage}\hfill
    \begin{minipage}{0.55\textwidth}
        \centering
        \includegraphics[width=\textwidth]{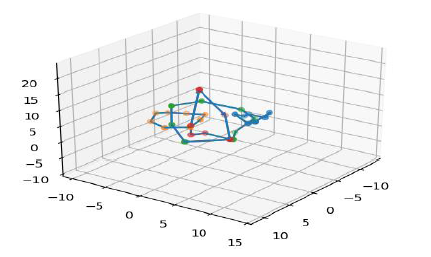}
        \caption{Removing one loop reduces the number of links from three to one.}
        \label{fig:image2}
    \end{minipage}
\end{figure}

The data points have been manually selected to create a structure with four interlinked loops. This example is kept simple for ease of understanding, though the method can be extended to more complex structures with additional loops. We can consider these loops as magnetic field lines evolving over time, with the depicted structure representing one instance in their evolution. At this instance, there are three distinct links present, which we attempt to count using our algorithm. A subroutine within the algorithm calculates the linking number for any given pair of loops. By extending this algorithm to loop over all possible pairs within the structure, we can determine the total number of links at a particular moment. The output is shown below.\\

\begin{lstlisting}
# Output Figure 18
The number of links identified is 3
\end{lstlisting}

Next, we remove one loop, as shown in the Figure 19 above, which can be thought of as representing the breaking of magnetic field lines during their evolution. This removal reduces the number of links from three to one, a change accurately captured by the algorithm.

\begin{lstlisting}
# Output Figure 19
The number of links identified is 1
\end{lstlisting}

This approach allows us to track how the number of links in a system changes over time, revealing whether the count increases or decreases at different time instances.

\section{Conclusion}
The algorithm has proven effective in successfully distinguishing between two different links. It performs particularly well when working with a smaller number of data points. In such cases, the algorithm operates efficiently, providing quick and accurate results. However, it is important to acknowledge that as the number of points increases, the computational time required by the algorithm also grows. This suggests that the algorithm’s performance is dependent on the scale of the input data. As the dataset becomes larger, the time complexity of the algorithm increases, resulting in progressively longer processing times. While the algorithm remains robust and accurate in distinguishing between links, its efficiency diminishes with larger datasets due to the increased computational load.

To address this challenge, the algorithm can be optimized through parallelization. By dividing the workload across multiple processors, the computational time can be significantly reduced, allowing the algorithm to handle larger and more complex datasets with greater efficiency. Parallelization would be particularly beneficial for analyzing extensive data sets, such as those encountered in fields like plasma physics or astrophysics.

Once the necessary modifications are implemented, we plan to extend our analysis to larger datasets of magnetic field intensities. A particularly exciting avenue of research involves astronomical data collected by various space missions. For example, the magnetic field intensity data recorded by NASA's Parker Solar Probe offers a valuable resource. By analyzing such data, we can trace the magnetic field lines over time and apply our algorithm to investigate how complex linkages of magnetic field lines form, evolve, and break apart. This study can also account for the viscosity of the medium, providing a deeper understanding of how magnetic reconnection events occur in space. The ability to trace and analyze these evolving magnetic structures is crucial for understanding phenomena such as solar flares and geomagnetic storms, as well as other processes driven by magnetic reconnection in astrophysical plasmas.  

\section*{Acknowledgment}
Our software `LINKAGE' is available in GitHub (https://rupakmukherjee.github.io/Ratul-Chakraborty-22UMPY15/) under MIT License as open-source software. All work has been performed in Sikkim University `Brahmagupta' HPC facility. One of the authors Rupak Mukherjee acknowledges IUCAA visiting associateship program for their kind support.

\bibliographystyle{unsrt}
\bibliography{biblio}

\end{document}